\documentclass{egpubl}
\usepackage{vcbm2026s}

 \WsShortPaper      % uncomment for Short Paper contribution
\usepackage[T1]{fontenc}
\usepackage{dfadobe}  

\usepackage{cite}  % comment out for biblatex with backend=biber
\BibtexOrBiblatex
\electronicVersion
\PrintedOrElectronic
\ifpdf \usepackage[pdftex]{graphicx} \pdfcompresslevel=9
\else \usepackage[dvips]{graphicx} \fi

\usepackage{egweblnk} 
\usepackage{amsmath}
\usepackage{amssymb}
\usepackage{booktabs}
\usepackage{multirow}
\usepackage{framed}
\usepackage{xcolor}
\title[MultiPASE]%
      {MultiPASE: Shape-Aware Comparative 3D Genome Analysis}

\author[N. Nipu et al.]
{\parbox{\textwidth}{\centering
        N.\,Nipu\thanks{mnipu2@uic.edu}$^{1}$\orcid{0009-0006-4602-0359}
        ,
        G.\,Johnson$^{2}$\orcid{0000-0002-4456-6909}
        ,
        C.\,Roth$^{3}$\orcid{0000-0003-4934-6160}
        ,
        D.\,Rogers$^{3}$\orcid{/0000-0002-4495-6026}
        ,
        S.\,R.\,Starkenburg$^{3}$\orcid{0000-0002-8564-1274}
        ,
        C.\,R.\,Steadman$^{3}$\orcid{0000-0002-9672-5529}
        and G.\,E.\,Marai$^{1}$\orcid{0000-0002-7212-9669}
        }
        \\
{\parbox{\textwidth}{\centering
        $^1$University of Illinois Chicago, USA\\
        $^2$National Laboratory of the Rockies, USA\\
        $^3$Los Alamos National Laboratory, USA
        }
}
}

\begin{document}

% uncomment for using teaser
% \teaser{
%  \includegraphics[width=0.9\linewidth]{eg_new}
%  \centering
%   \caption{New EG Logo}
% \label{fig:teaser}
%}

\maketitle
%-------------------------------------------------------------------------
\begin{abstract}
   3D genome structures reconstructed from Hi-C data provide valuable insights into chromatin organization, but comparing them across conditions and time points remains challenging because of noise, variability, and arbitrary orientation. State-of-the-art comparison approaches often rely on contact-matrix-based summaries, which do not preserve the spatial geometry of the reconstructed structures and make integration with gene expression and chromatin accessibility difficult. There is therefore a need for robust structural alignment and spatial abstraction methods that support integrated analysis across conditions and time points. We present MultiPASE, a computational pipeline for comparative analysis of reconstructed 3D genome structures represented as gene-level spatial point clouds. MultiPASE aligns point clouds into a shared coordinate frame using coarse-to-fine rigid registration with PCA-based axis disambiguation and robust ICP refinement, then summarizes spatial organization using two complementary shape abstraction methods that produce stable, nested, multi-scale representations from dense core regions to broader structural extent. Using Hi-C-derived reconstructions from viral infection experiments, we show how the approach reveals condition-dependent structural changes that are difficult to interpret from raw 3D structures alone. By integrating spatial structure with gene expression and chromatin accessibility, MultiPASE supports interactive analysis of spatial organization and gene relocation patterns. A quantitative validation, reported in the Supplementary Materials, shows that the proposed shape abstractions preserve non-convex features, remain stable across samples, and that the alignment pipeline is necessary for meaningful comparison. A usage case and a usability evaluation with eight participants demonstrate the pipeline's effectiveness for comparative structural analysis.
%-------------------------------------------------------------------------
%  ACM CCS 1998
%  (see https://www.acm.org/publications/computing-classification-system/1998)
% \begin{classification} % according to https://www.acm.org/publications/computing-classification-system/1998
% \CCScat{Computer Graphics}{I.3.3}{Picture/Image Generation}{Line and curve generation}
% \end{classification}
%-------------------------------------------------------------------------
%  ACM CCS 2012
   % (see https://www.acm.org/publications/class-2012)
%The tool at \url{http://dl.acm.org/ccs.cfm} can be used to generate
% CCS codes.
%Example:
% \begin{CCSXML}
% <ccs2012>
% <concept>
% <concept_id>10010147.10010371.10010352.10010381</concept_id>
% <concept_desc>Computing methodologies~Collision detection</concept_desc>
% <concept_significance>300</concept_significance>
% </concept>
% <concept>
% <concept_id>10010583.10010588.10010559</concept_id>
% <concept_desc>Hardware~Sensors and actuators</concept_desc>
% <concept_significance>300</concept_significance>
% </concept>
% <concept>
% <concept_id>10010583.10010584.10010587</concept_id>
% <concept_desc>Hardware~PCB design and layout</concept_desc>
% <concept_significance>100</concept_significance>
% </concept>
% </ccs2012>
% \end{CCSXML}

% \ccsdesc[300]{Computing methodologies~Collision detection}
% \ccsdesc[300]{Hardware~Sensors and actuators}
% \ccsdesc[100]{Hardware~PCB design and layout}

\printccsdesc   
\end{abstract}  
%-------------------------------------------------------------------------

\section{Introduction}
The spatial organization of the genome is integral to the regulation of gene expression, chromatin accessibility, and cellular responses to perturbations~\cite{Dekker2013Exploring,Bonev2016Organization}. Recent advances in chromosome conformation capture techniques, such as Hi-C, has enabled the reconstruction of three-dimensional (3D) genome structures across conditions and time~\cite{Aiden2009Comprehensive,Imakaev2012Iterative}. Hi-C measures interaction frequencies between genomic regions, which are commonly represented as contact matrices~\cite{Aiden2009Comprehensive,Belton2012HiC}. These reconstructions have generated growing interest in understanding how genome architecture changes during processes such as viral infection~\cite{Wang2023Sars}, cellular differentiation~\cite{Dixon2015Chromatin}, and stress responses~\cite{Vihervaara2017Transcriptional}. This raises the question of how reconstructed structures can be compared across conditions and whether such comparisons reveal patterns that are difficult to identify from individual structures.

However, comparing reconstructed 3D genome structures across conditions is challenging due to noise and sensitivity to experimental variability and artifacts~\cite{Trussart2015Assessing,Serra2017Automatic}. Each gene also has a 3D position and attributes such as expression and accessibility across conditions and time. Existing methods often focus on summaries from Hi-C contact matrices rather than the actual reconstructed structures~\cite{Aiden2009Comprehensive,Rao2014A3dMap,Bonev2016Organization}, which lose spatial geometry. While some work allows for visualization of 3D structures~\cite{Trussart2015Assessing,Serra2017Automatic}, they provide limited support for alignment, shape-based abstraction, and systematic comparisons across conditions and time points. A further challenge arises when integrating structural data with functional measurements. While gene expression and chromatin accessibility are studied alongside genome organization~\cite{Schoenfelder2019Long,Ulianov2016Active,Paulsen2019Longrange}, they are often evaluated separately from global 3D geometry. Mapping genes onto reconstructed 3D structures enables exploration of spatial reorganization in relation to changes in expression and accessibility. Such analysis requires methods that align structures, minimize reconstruction noise, and create comparable spatial summaries across conditions and time points.

From a data visualization perspective, this problem relates to ensemble visualization and, more specifically, to second-generation ensemble data~\cite{nipu2026neurosurgery}. Unlike traditional ensemble data, where spatial representations are predefined or can be obtained directly from individual members~\cite{wang2018visualization}, second-generation ensembles require data transformation to derive the representations needed for comparison~\cite{nipu2023visual,luciani2018details}. In visualization, data transformation refers to visualization algorithms and techniques that transform data from one form to another to support the intended visual representation~\cite{ieeevis2026areamodel,eurovis2026fullpapers}. Reconstructed 3D genome structures fit this setting because they are noisy, unregistered point clouds with arbitrary orientation and variable point counts. Meaningful comparison therefore requires transformation through alignment, abstraction, and aggregation before the structures can be analyzed together.

In this work, we present Multi-Point Alignment and Shape Extraction (MultiPASE), a publicly available Python package for comparative analysis of reconstructed 3D genome structures. Using Hi-C-derived reconstructions, we map genes into 3D space and integrate gene-level expression and chromatin accessibility measurements. MultiPASE aligns reconstructed point clouds into a shared coordinate frame and transforms them into comparable shape-based representations. Density-based representations help reduce sensitivity to local variation while preserving broader spatial organization~\cite{Silverman1986Density,Ester1996Density}. We also present a web-based visual analytics frontend for exploring aligned structures, shape abstractions, gene expression, and chromatin accessibility across conditions and time points. We demonstrate the approach using chromatin rearrangements associated with viral infection and evaluate it through quantitative validation, a usage case, and a usability study.

The contributions of this work are as follows:
(1) MultiPASE, a computational pipeline for aligning reconstructed 3D genome point clouds into a shared coordinate frame using PCA-based axis disambiguation and rigid ICP refinement.
(2) Two complementary shape abstraction methods, highest-density-region (HDR) and point-fraction (PF) silhouettes, with quantitative comparison metrics for multi-scale structural analysis.
(3) An interactive visual analytics tool with coordinated views for exploring aligned 3D genome structures, 2D shape abstractions, and gene expression and chromatin accessibility correlation across conditions and time points. (4) Quantitative validation, a usage case, and a usability study reported in the Supplementary Materials that evaluate the approach for comparative genome structure analysis.

\section{Related Work}
\noindent\textbf{Ensemble Visualization. }Ensemble visualization addresses collections of related spatiotemporal datasets generated under varying initial conditions, parameters, or configurations, and supports their comparison through statistical summaries, aggregation, and spatial representations~\cite{wang2018visualization,bonneau2014overview}. Methods such as contour and curve boxplots summarize variability across spatial structures~\cite{whitaker2013contour,mirzargar2014curve}, while probabilistic approaches represent uncertainty in spatial regions and surfaces~\cite{poethkow2011probabilistic,pfaffelmoser2011visualizing}. More recent work addresses second-generation ensembles, where spatial representations must first be derived through transformation and aggregation~\cite{luciani2018details,nipu2023visual,nipu2026neurosurgery}. Reconstructed 3D genome structures share this setting, but consist of noisy, unregistered point clouds with arbitrary orientation and variable point counts. MultiPASE addresses this setting through alignment and multi-scale shape abstraction. 

\noindent\textbf{3D Genome Visualization and Comparison. }Most genome visualization systems focus on genomic sequences~\cite{aurisano2015bactogenie,luciani2014fixingtim}, Hi-C contact matrices, or individual reconstructed structures~\cite{chen2026chrompolymerdb}. HiGlass~\cite{Kerpedjiev2018HiGlass}, Juicebox~\cite{Durand2016Juicebox}, and HiPiler~\cite{lekschas2018hipiler} support exploration of interaction matrices, while CSynth~\cite{Todd2021CSynth} and GenomeFlow~\cite{Trieu2019GenomeFlow} support interactive inspection of reconstructed 3D structures. Other systems integrate functional annotations with genome structure~\cite{zhu2022nucleome,li2022washu,Rogers2026From}. Computational comparisons of genome organization also commonly rely on contact-matrix-derived summaries~\cite{Aiden2009Comprehensive,Rao2014A3dMap,Bonev2016Organization}. These approaches provide important insights into genome organization but offer limited support for alignment and shape-based comparisons of reconstructed 3D structures across conditions and time points. MultiPASE operates directly on reconstructed gene-level point clouds and connects spatial comparison with gene expression and chromatin accessibility.

\noindent\textbf{Point Cloud Registration and Shape Analysis. }
Comparing reconstructed genome structures requires aligning point clouds that do not share a common coordinate system. Rigid registration methods such as ICP~\cite{besl1992method,Rusinkiewicz2001EfficientICP} and Procrustes analysis~\cite{Schonemann1966Generalized} estimate transformations between point sets. MultiPASE uses PCA-based initialization followed by trimmed rigid ICP to remove arbitrary rotation and translation while preserving the structural differences that are later compared. Density-based methods such as KDE~\cite{tsin2004correlation}, GMM~\cite{jian2011robust}, and CPD~\cite{Myronenko2010CPD} can improve registration of noisy point sets using smoother distributional objectives, but their alignment depends on choices such as bandwidth, mixture representation, scale, and optimization settings. MultiPASE instead separates alignment from shape abstraction, applying smoothing later through HDR and PF. Non-rigid registration is avoided because it could absorb condition-dependent structural differences that we aim to preserve. Learning-based approaches such as GeoTransformer~\cite{Qin2022CVPR}, MADBD~\cite{zhang2026madbd}, and C-GenReg~\cite{Haitman2026CVPR} require learned representations, training data, or image appearance that are not available for the reconstructed genome point clouds considered here.

For shape analysis, shape contexts~\cite{Belongie2002ShapeContexts}, spin images~\cite{Johnson1999SpinImages}, and point feature histograms~\cite{Rusu2009FPFH} describe local point-cloud geometry for matching and recognition. Alpha shapes~\cite{Edelsbrunner1983Shape,Edelsbrunner1994AlphaShapes} capture non-convex boundaries, but depend on a scale parameter that can produce fragmented or inflated shapes across noisy point clouds. MultiPASE instead uses HDR and PF to derive nested, multi-scale shape representations for comparison across conditions and time points.

\section{Methods}
\begin{figure}[t]
    \centering
    \includegraphics[
        width=\linewidth,
        alt={Diagram of the system workflow showing raw Hi-C, gene annotation, expression, and accessibility data processed through MultiPASE to produce aligned gene coordinates, shape abstractions, and temporal agreement classes for use in an interactive visual frontend.}
    ]{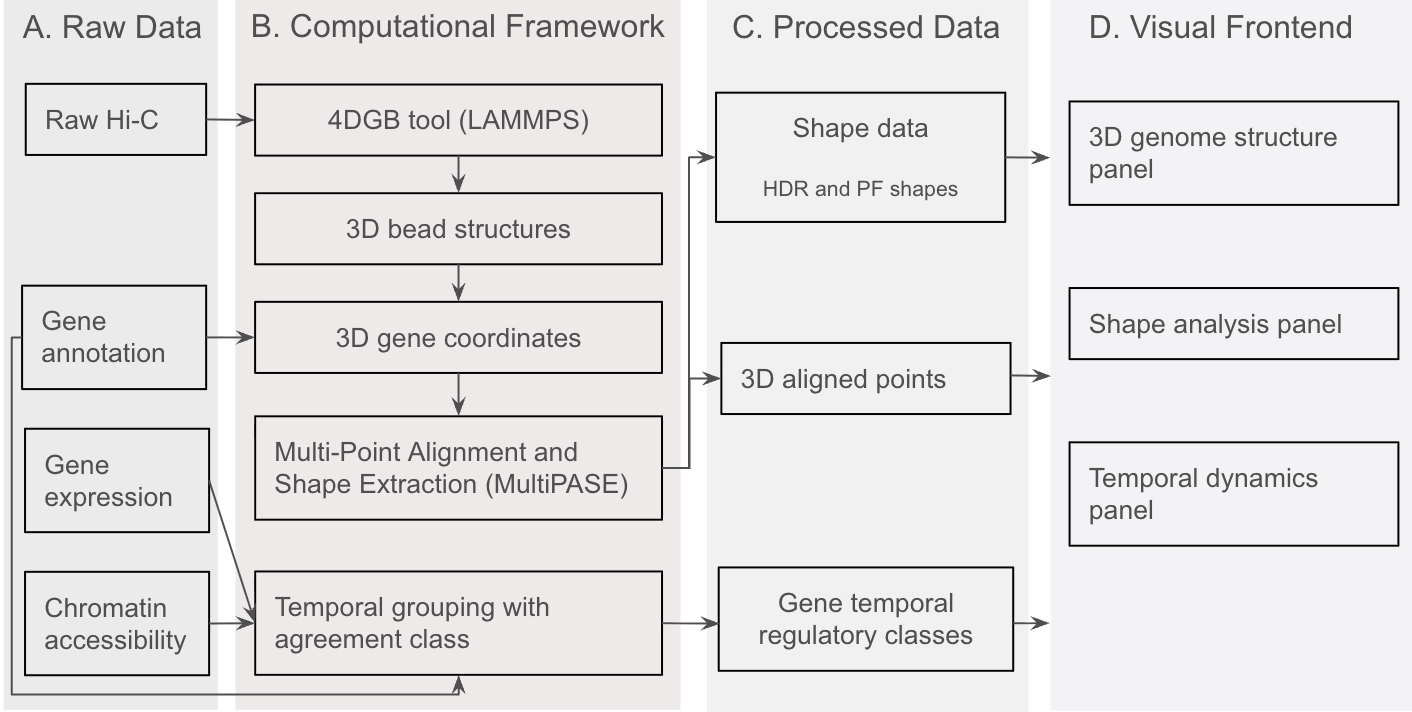}
    \caption{System architecture. (A) Hi-C data are used to reconstruct 3D genome structures and derive gene-level coordinates. (B) MultiPASE aligns these gene point clouds across conditions and extracts shape-based summaries, while expression and accessibility data are grouped over time to define regulatory agreement classes. (C-D) These processed outputs support interactive exploration in the visual frontend.}
    \label{fig:system}
\end{figure}
\subsection{Input Data and Preprocessing}
All analyses were conducted using Vero Hi-C-based 3D genome reconstructions from African green monkey cells under uninfected and vaccinia virus infected conditions~\cite{Venu2024multiOmics}. \textcolor{black}{The source study collected uninfected and infected samples at 12, 18, and 24 hours post-infection, with two biological replicates per condition and time point. In our analysis, we used one Hi-C replicate per condition-time combination to generate chromosome-level 3D reconstructions. We analyzed 24 chromosomes for which reconstruction outputs and matched gene-level measurements were available.} Gene annotations, gene expression measurements, and chromatin accessibility data were matched using consistent gene identifiers, enabling integration of spatial, transcriptional, and regulatory information at the gene level across conditions and time points.

\noindent\textbf{Hi-C-based 3D Genome Reconstruction. }Hi-C data were processed using the 4DGB workflow tool~\cite{Rogers2026From} to extract 3D structure files for each condition and time point. We obtained bead-based representations at a resolution of 100 kb using LAMMPS Molecular Dynamics Simulation~\cite{Thompson2022Lammps, Anna2021Four}, which is integrated within the tool. Each bead corresponds to a contiguous 100 kb genomic locus. \textcolor{black}{For each condition and time point, we generated chromosome-level 3D coordinate files for the 24 analyzed chromosomes.} These reconstructed bead coordinates provide the structural input for subsequent analysis (Fig.~\ref{fig:system}.A-B).

\noindent\textbf{Gene Mapping to 3D Coordinates and Integration of Expression and Accessibility Data. }
Genes were mapped to chromosome-level 3D structures using genomic annotations. Because the reconstructions have 100 kb resolution, gene positions were estimated by linear interpolation between neighboring bead coordinates, and each gene was represented by its interpolated midpoint in 3D space. The spatial precision of these positions is therefore limited by the reconstruction resolution. Gene expression differences between infected and uninfected conditions were integrated with these positions. Chromatin accessibility was obtained from ATAC-seq narrowPeak files using the peak closest to each gene's transcription start site (TSS), and accessibility differences were computed between infected and uninfected conditions at each time point.

\subsection{Task Analysis}
Through discussions with domain experts in viral infection genomics, we identified six analytical tasks that guided the design of the computational pipeline and visual interface.

\begin{itemize}
\itemsep0em \vspace{-0.3em}
\item \vspace{-0.3em} T1. Align reconstructed 3D genome structures across conditions and time points into a shared coordinate frame to remove arbitrary translation and orientation.
\item \vspace{-0.3em} T2. Summarize gene-level point clouds as stable shape representations that reduce noise while preserving global structure.
\item \vspace{-0.3em} T3. Quantify structural differences across conditions and time points at multiple abstraction levels, from dense gene regions to broader structural extent.
\item \vspace{-0.3em} T4. Identify when and where structural reorganization occurs across time points and projection planes.
\item \vspace{-0.3em} T5. Examine how genes with different regulatory behaviors are spatially distributed and reorganized under infection.
\item \vspace{-0.3em} T6. Relate gene spatial changes to expression and chromatin accessibility changes.
\end{itemize}
\vspace{-0.4em}

Tasks T1-T3 motivated the design of MultiPASE, while T4-T6 guided the coordinated visual interface. The usability evaluation tasks were derived from these analytical needs.

\subsection{Temporal Grouping and Agreement Classes}
To analyze gene expression and chromatin accessibility dynamics over time, we categorized genes based on their temporal response patterns across the three time points. For each gene, we first labeled expression and TSS-based accessibility differences as increased, decreased, or unchanged at each time point. Larger positive and negative values were labeled as increases or decreases, while minor changes were considered to be no change. Expression and accessibility directions were then compared at each time point to determine whether they changed together, in opposite directions, or independently. Based on these temporal patterns, we classified genes into seven categories: both increase, both decrease, expression-only, accessibility-only, mixed, conflict, and stable. These agreement classes were used as categorical labels in subsequent analyses to examine how genes with similar temporal behavior are organized within the 3D genome.

\subsection{MultiPASE: Alignment and Shape-based Abstraction of 3D Genome Structures}
\textcolor{black}{Reconstructed 3D genome structures provide detailed spatial information, but comparing them across conditions and time points is challenging because each reconstruction may differ in orientation, scale, and local structural variability. To address these challenges, we introduce MultiPASE (Multi-Point Alignment and Shape Extraction), a computational pipeline that aligns gene-level point clouds into a shared coordinate frame (T1), extracts density-based highest-density region (HDR) and point-fraction (PF) shape abstractions (T2), and quantifies structural differences using overlap and boundary-distance metrics (T3) Fig.~\ref{fig:system}.B-C).}

For each condition or time point, the input to MultiPASE is a set of 3D points where each point represents the spatial position of a gene mapped from Hi-C-derived reconstructions.

\subsubsection{Structural Alignment}
Reconstructed 3D genome structures have arbitrary position and orientation, which can obscure structural comparison. MultiPASE therefore aligns all point sets to a shared reference using three stages: centering, PCA-based pre-alignment with axis disambiguation, and rigid ICP refinement. Each point set is first translated to the origin by subtracting its centroid. We restrict alignment to rotation and translation because non-rigid deformation could absorb structural differences that we aim to preserve. For each chromosome, all structures are aligned to the earliest uninfected time point.

\noindent\textbf{PCA-based Pre-alignment. }
PCA estimates the dominant spatial axes of each centered point cloud and provides an initial orientation for ICP~\cite{jolliffe2011principal,besl1992method}. Because PCA axes can be permuted or sign-flipped, direct PCA alignment can still produce mismatched orientations. We therefore evaluate all six axis permutations and eight sign combinations. Of the resulting 48 candidates, the 24 that introduce reflections are discarded. The remaining proper rotations are evaluated using symmetric nearest-neighbor RMSE, and the lowest-error rotation is selected:

\begin{equation}
\begin{aligned}
\mathrm{RMSE}(A,B) = \Biggl[ \frac{1}{|A|+|B|} \biggl( &\sum_{b\in B} d(bR^\top,A)^2 \\
&+ \sum_{a\in A} d(a,BR^\top)^2 \biggr) \Biggr]^{1/2}
\end{aligned}
\end{equation}

where $d(x,P)$ is the Euclidean distance from point $x$ to its nearest neighbor in point set $P$.

\noindent\textbf{ICP Refinement. }
The PCA-aligned point cloud is refined using rigid ICP~\cite{besl1992method}. At each iteration, nearest-neighbor correspondences are computed and the 10\% largest residuals are removed to reduce the influence of noise and outliers. The remaining correspondences are used to estimate the rigid transformation with the Kabsch algorithm~\cite{kabsch1976solution}. The transformation is applied iteratively for a fixed number of iterations (default 30). For larger datasets, MultiPASE can subsample each point set before ICP (default maximum 50,000 points); in our genome dataset, chromosome-level point counts were below this threshold, so all points were used.

\subsubsection{Shared Scaling and Grid Construction}
After alignment, all point sets are normalized to a shared scale so that projections and shape abstractions from different conditions occupy the same spatial extent. A common spatial grid is then constructed with identical bin boundaries for all structures. The grid resolution is configurable, with 160 bins along the longest axis by default. The aligned points are projected onto the XY, YZ, and XZ planes using this shared grid, allowing density estimates and extracted shapes to be compared at corresponding spatial locations.

\begin{figure*}[t]
    \centering
    \includegraphics[
        width=1\linewidth,
        alt={Diagram of the GenomeVis interface with a navigation bar, a 3D genome structure view comparing conditions and time points, and a shape analysis panel showing 2D projections of gene distributions.}
    ]{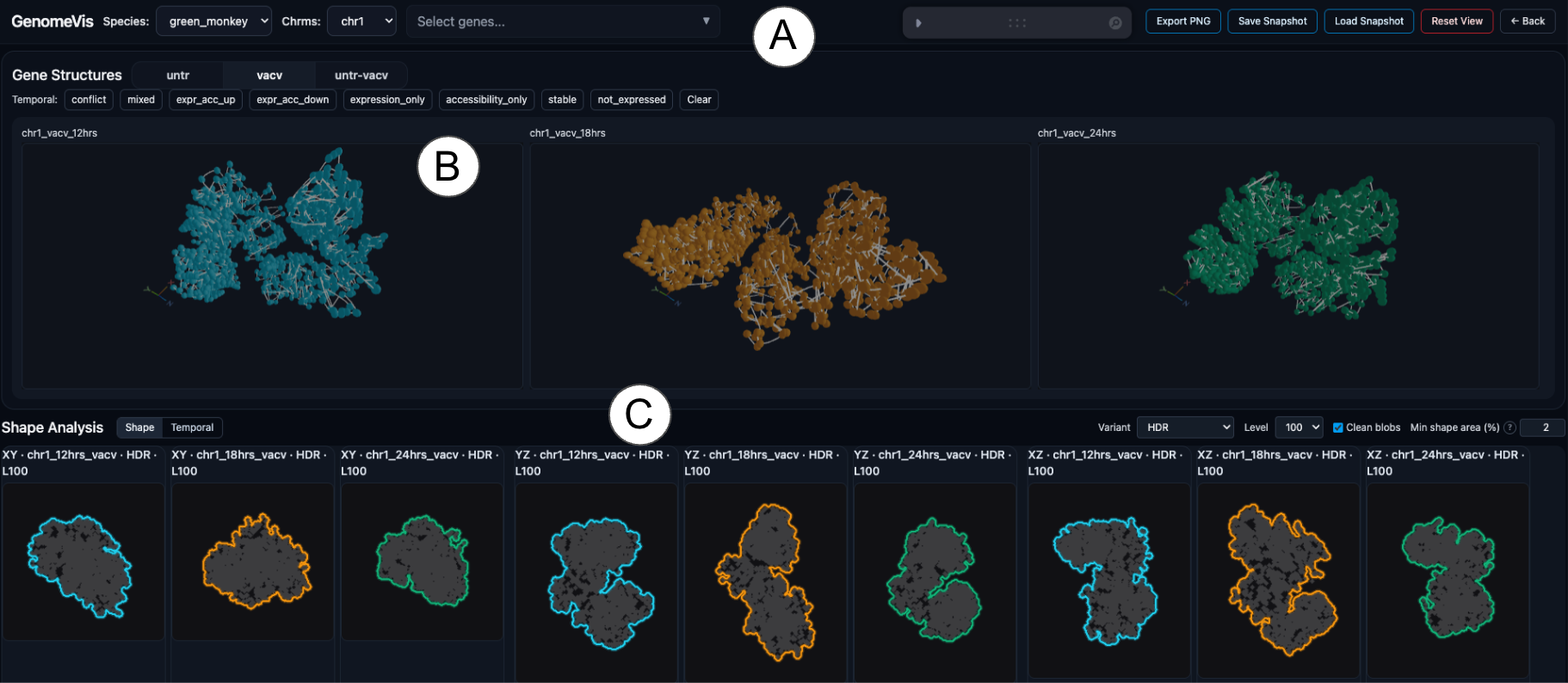}
    \caption{\textcolor{black}{GenomeVis, the visual frontend for MultiPASE-based 3D genome structure reorganization analysis.} (A) The navigation panel supports selecting species and chromosomes, managing gene selections, and saving or exporting analysis snapshots. (B) The 3D genome structure panel displays aligned genome structures across conditions and time points, enabling side-by-side spatial comparison. \textcolor{black}{(C) The shape analysis panel shows HDR/PF shape abstractions computed from the same aligned gene-level point clouds shown in the 3D structure panel. The 3D gene coordinates are projected onto the XY, YZ, and XZ planes, with three panels per projection corresponding to the three time points.}}
    \label{fig:frontend}
\end{figure*}

\subsubsection{Density-based Shape Abstraction}
MultiPASE combines aligned 3D views with projected 2D shape abstractions. The 3D views preserve spatial context, while the XY, YZ, and XZ projections provide fixed orientations for direct visual and quantitative comparison across conditions and time points. These projections reduce viewpoint, occlusion, and depth-perception effects, but they collapse depth information and are therefore used as complementary summaries rather than replacements for the 3D structures. The next challenge is to summarize each projection as a comparable shape. Convex hulls can overestimate spatial extent by filling concavities and expanding toward outliers. Alpha shapes preserve concavities, but depend on a scale parameter that can be difficult to apply consistently across noisy structures. Large values can inflate the shape, while small values can fragment it. Defining projection planes separately from the local PCA axes of each structure would also make the view frame depend on the individual point cloud. MultiPASE instead aligns all structures into a shared coordinate frame and projects them onto the same canonical planes. It then derives shapes using highest-density regions (HDR) and point-fraction (PF) representations. 

For HDR, MultiPASE uses bootstrap aggregation to obtain stable density estimates~\cite{tibshirani1993introduction}. For each condition, time point, and projection plane, it generates $B$ bootstrap resamples of the projected point set, with $B=256$ by default. Each resample contains the same number of points as the original set and is sampled with replacement. The resample is rasterized onto the shared grid and smoothed with a Gaussian kernel~\cite{Silverman1986Density}. The smoothed density maps are then averaged:

\begin{equation}
D(u) =
\frac{1}{B}\sum_{b=1}^{B}
\mathcal{G}_\sigma\!\left(
\mathrm{Hist}\left(\{u^{(b)}_i\}\right)
\right)
\end{equation}

where $u$ denotes grid locations, $G_{\sigma}$ is Gaussian smoothing with bandwidth $\sigma$, and $u_i^{(b)}$ is the $i$th projected point in bootstrap sample $b$. The default bandwidth is $\sigma=1.2$ pixels. Before shape extraction, density values below 0.2\% of the maximum are set to zero to reduce the influence of faint density from distant outliers. 

HDRs are then extracted at different density-mass levels~\cite{hyndman1996computing}. For a target mass fraction $m \in (0,1]$, MultiPASE selects a threshold $\tau_m$ such that the retained region contains approximately fraction $m$ of the total density: 

\begin{equation} 
\sum_{u:D(u)\geq\tau_m} D(u) \approx m \sum_u D(u)
\end{equation}

In practice, density values are sorted from highest to lowest and accumulated until the target mass is reached. Grid locations at or above the resulting threshold form the HDR mask, and the outer contour of the largest connected component defines the HDR shape. Varying $m$ produces nested shapes, where lower mass fractions represent dense core regions and higher fractions capture broader structural extent.

\subsubsection{Point-fraction Shape Abstraction}

To complement HDR, MultiPASE extracts point-fraction (PF) shapes that remain directly linked to individual gene locations. While HDR derives shapes from a bootstrap-averaged density field, PF ranks the projected genes by their local point density and retains the densest subset. This provides a complementary view of spatial organization at the point level.

For each condition, time point, and projection plane, the 3D coordinates are projected to 2D, yielding point locations $\{u_i\}_{i=1}^{N}$. Local point density is estimated using Gaussian kernel density estimation (KDE)~\cite{Silverman1986Density}:

\begin{equation}
\hat{f}(u) =
\frac{1}{Nh^2}
\sum_{j=1}^{N}
\exp\left(
-\frac{\|u-u_j\|^2}{2h^2}
\right),
\end{equation}

where $h$ is the kernel bandwidth. When not specified by the user, $h$ is set to half the median pairwise distance estimated from a subsample of points. This provides a data-dependent scale for estimating local point concentration without excessive smoothing.

Each point is assigned a density score and ranked from highest to lowest density. For a target fraction $f \in (0,1]$, the top $fN$ points are retained. Smaller fractions emphasize the densest regions, while larger fractions include more peripheral points and capture broader structural extent. Because points are added in decreasing order of density, PF shapes are nested across fraction levels. Low-density outliers are also excluded at stricter fractions. The retained points are then rasterized onto the shared projection grid using disks with a configurable radius of 2 pixels by default. The resulting binary mask is refined using morphological closing and opening to fill small gaps and remove isolated regions~\cite{SerraImage1983}. The largest connected region is retained and interior holes are filled. Finally, its outer contour is extracted using a marching-squares procedure~\cite{Lorensen1987Marching}, producing the PF shape for the selected projection and fraction level.

Together, HDR and PF provide complementary multi-scale representations. HDR summarizes spatial organization through bootstrap-stabilized density mass, while PF preserves a direct connection to density-ranked gene locations. Both produce nested shapes on the same shared grid, enabling consistent comparison across conditions and time points.

\subsubsection{Quantitative Comparison Metrics}
MultiPASE compares matched shape abstractions separately for each projection plane and abstraction level using Intersection-over-Union (IoU), Hausdorff distance~\cite{Huttenlocher1993Comparing}, and mean nearest-neighbor (meanNN) boundary distance. IoU measures overall shape overlap, Hausdorff distance captures the largest boundary separation, and meanNN summarizes typical boundary separation. Metrics are computed across HDR mass and PF fraction levels such as 100\%, 95\%, 90\%, 80\%, 60\%, and 50\%, ranging from dense core regions to broader structural extent.

\subsection{System Design and Implementation}
Our system includes a computational backend and an interactive visual frontend (Fig.~\ref{fig:system}). It processes raw data to extract 3D genome structures along with gene expression and chromatin accessibility measurements. MultiPASE aligns structures across conditions and time points using the earliest uninfected time point as a reference (Fig.~\ref{fig:system}.B), providing a shared coordinate frame for comparison. After alignment, MultiPASE extracts shape representations and computes metrics summarizing structural differences. Outputs include aligned 3D point sets, shape representations, background masks, and comparison metrics (Fig.~\ref{fig:system}.C), which facilitate interactive exploration in the visual frontend (Fig.~\ref{fig:system}.D). The backend and MultiPASE are implemented in Python, while the frontend uses React, D3.js, and Three.js. MultiPASE is available at \url{https://github.com/nafiul-nipu/MultiPASE.git}, with the visual frontend at \url{https://nafiul-nipu.github.io/GenomeVisProject}. The complete system can be found at \url{https://github.com/nafiul-nipu/GenomeVisProject.git}.

\subsection{Data Visualization Design}
The visual front end, designed through an activity-centered-design approach~\cite{marai2017activity}, comprises three coordinated components which couple the 3D structures familiar to the biologists~\cite{marai2015visual} with the new shape abstraction: 1) the 3D genome structure panel aids in exploring spatial genome organization across time and conditions (T4); 2) the shape analysis panel assists in examining structural evolution and gene relocation through shape abstraction (T4, T5), and 3) the temporal dynamics panel helps in comparing gene expression and accessibility changes across genes (T6). The top navigation panel supports exploration across species, chromosomes, and individual genes (Fig.~\ref{fig:frontend}.A).  This panel also provides functionality to export visualizations as images and save complete session states for reproducible sharing of analytical findings. 

\noindent\textbf{1) 3D Genome Structure Panel:} The 3D genome structure panel provides a multi-tabbed view for comparing uninfected and infected genome structures at all timepoints, including direct comparisons(T4) (Fig.~\ref{fig:frontend}.B). Genes are displayed as points, color-coded by timepoint, with a tube representing the chromatin structure. Users can interactively zoom, pan, and rotate, with changes mirrored across views for easier comparison. The panel allows filtering of genes by temporal agreement class, hiding non-selected genes and highlighting remaining ones in class-specific colors. Hovering over a gene reveals its spatial context in the shape analysis panel.

\begin{figure}[h]
    \centering
    \includegraphics[
        width=\linewidth,
        alt={Sequence of shape-based summaries in the YZ projection showing how gene spatial concentration changes over time under infection, including dispersion at the intermediate time point and partial recovery at the final time point.}
    ]{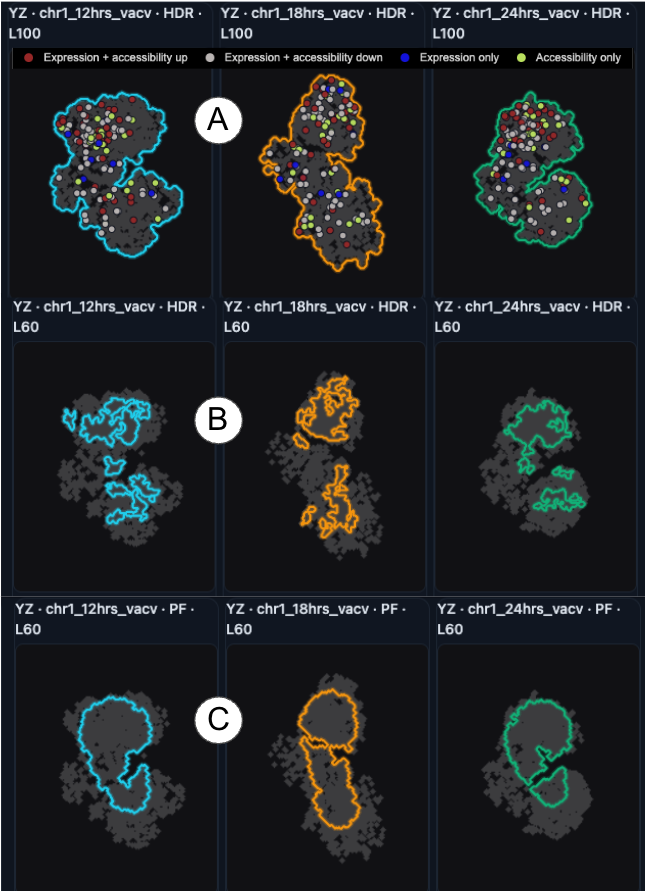}
    \caption{Usage case demonstrating gene relocation and concentration changes in the YZ projection for chromosome 1 under infection. (A) Gene relocation patterns across time show genes concentrated in the upper region at 12 h, becoming more dispersed at 18 hrs, and partially returning by 24 hrs. (B) The 60\% highest-density region (HDR) highlights an expansion of core gene concentration at 18 hrs, followed by contraction at 24 hrs. (C) The 60\% point-fraction (PF) representation shows a similar temporal pattern while covering a broader spatial area.}
    \label{fig:concentration}
\end{figure}

\noindent\textbf{2) Shape Analysis Panel:} The shape analysis panel allows users to explore gene distribution and structural evolution through 2D projections (XY, YZ, XZ) (T4, T5) (Fig.~\ref{fig:frontend}.C). It shows shape summaries overlaid on a background mask, displayed either across timepoints for a condition or across conditions for a timepoint. Users can select HDR or PF representations to analyze patterns, with colors consistent with the 3D genome structure panel for easy comparison. Abstraction levels can be adjusted to refine gene concentration, and noise filtering removes small isolated regions for clearer interpretation. Temporal class filtering highlights the spatial distribution of selected genes (Fig.~\ref{fig:concentration}.A). Hovering over a shape reveals the corresponding 3D structure, connecting abstract patterns to spatial configurations.

\begin{figure}[t]
    \centering
    \includegraphics[
        width=\linewidth,
        alt={Scatterplot of gene expression change versus chromatin accessibility change, paired with a gene-level panel showing expression and accessibility trends over time for selected genes.}
    ]{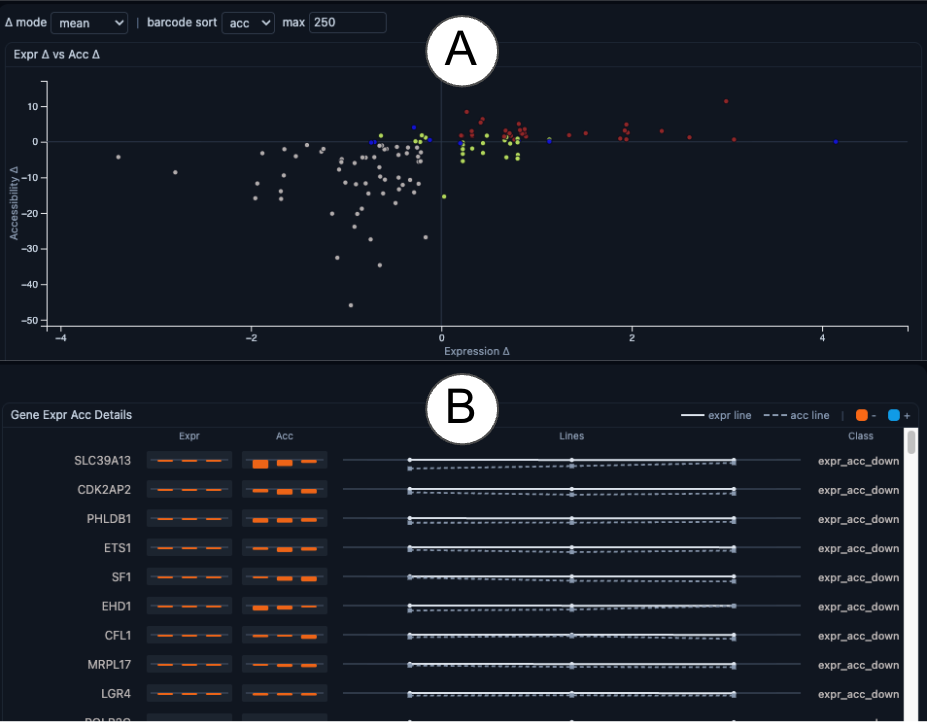}
    \caption{Temporal dynamics panel. (A) Scatterplot showing changes in gene expression and chromatin accessibility, with genes colored by their temporal agreement class. (B) Gene-level detail view showing expression and accessibility trends over time, allowing inspection of their relationship at the individual gene level.}
    \label{fig:temporal}
\end{figure}

\noindent\textbf{3) Temporal Dynamics Panel:} The temporal dynamics panel enables the analysis of gene expression and accessibility changes over time alongside spatial reorganization (T6). It features two coordinated views. The expression-accessibility scatterplot displays gene distribution based on expression and accessibility changes, with options to aggregate data by mean change, final timepoint, or peak values (Fig.~\ref{fig:temporal}.A). Genes are color-coded by temporal class, and hovering reveals specific details. Users can select subsets of genes to filter other panels. The second view shows gene-level details over time, with positive values in blue and negative in orange, including trajectory lines for expression (solid) and accessibility (dotted) (Fig.~\ref{fig:temporal}.B). Hovering provides additional temporal information, and selecting a gene highlights it across all panels and updates the gene selection dropdown.

\section{Evaluation and Results}
We evaluate MultiPASE through a usage case that examines whether aligned shape abstractions and coordinated views support comparison of reconstructed 3D genome structures across conditions and time points (T4-T6). Quantitative validation of the alignment and shape abstraction pipeline (T1-T3), together with the usability study, is reported in the Supplementary Materials.

\subsection{Usage Case Demonstration}
This usage case investigates structural reorganization of genome architecture and gene relocation over time (T4, T5), and examines how these spatial changes relate to gene expression and chromatin accessibility (T6). In particular, the analysis explores whether infection-associated spatial reorganization corresponds to changes in expression or accessibility, and whether expression changes consistently align with accessibility changes.

For this demonstration, we analyzed publicly available data from the African green monkey (Chlorocebus sabaeus)~\cite{Venu2024multiOmics}, focusing on chromosome 1 under uninfected (UNTR) and infected (VACV) conditions at 12hrs, 18hrs, and 24hrs using 3D genome structure analysis. The uninfected condition remained stable, while the infected condition showed largest structural deviations at 18hrs. Uninfected comparisons indicated higher similarity (e.g., IoU 0.592 to 0.601 and meanNN 2.86 to 4.15 for YZ at 95\%), whereas infected comparisons revealed reduced similarity at 18hrs (IoU 0.502, meanNN 5.24 from 12hrs to 18hrs in YZ). The YZ projection highlighted elongation at 18hrs followed by partial reversion at 24hrs (Fig.~\ref{fig:frontend}.B). These observations indicate that infection induced a transient structural reorganization of chromosome 1, and that the changes visible in the 3D view are consistently captured by shape-based abstractions.

We next examined how infection affected gene distribution. In the 60\% highest-density region, core gene occupancy expanded at 18 hours and contracted at 24 hours (Fig.~\ref{fig:concentration}.B). PF representations showed a similar pattern with a broader spatial extent (Fig.~\ref{fig:concentration}.C). At 18 hours, we observed lower IoU and higher meanNN compared to 12 and 24 hours, with values of 0.240 / 5.07 in YZ and 0.179 / 22.70 in XZ at 60\%. The XZ projection had particularly high meanNN values during infection, peaking at 55.03 from 18 to 24 hours, indicating significant spatial deformation.

Finally, we examined the relationship between gene expression and chromatin accessibility. Filtering by temporal regulatory classes, including expression-only, expression-accessibility up/down, and accessibility-only, showed that genes with large expression changes often exhibited minimal accessibility changes, whereas genes with higher accessibility changes tended to show smaller expression changes (Fig.~\ref{fig:temporal}.A). The gene-level detail view further showed diverse temporal behaviors, including cases where large accessibility changes occurred without corresponding expression changes (Fig.~\ref{fig:temporal}.B). This finding is consistent with prior analyses of the dataset~\cite{Venu2024multiOmics}. 

\section{Discussion and Conclusion}

MultiPASE combines alignment, multi-scale shape abstraction, quantitative comparison, and coordinated visual analysis for reconstructed 3D genome structures. It enables comparison across conditions and time points while reducing sensitivity to local point-level variation.

From a visualization perspective, reconstructed genome structures form a second-generation ensemble dataset in which useful spatial representations must be derived from noisy and variable instances. In the usage case, MultiPASE revealed a transient reorganization of chromosome 1 at 18 hrs, followed by partial return at 24 hrs. The shape views also showed changes in gene concentration that were difficult to interpret directly from the 3D structures, while the temporal views supported prior observations that large expression changes do not necessarily correspond to changes in chromatin accessibility~\cite{Venu2024multiOmics}. The visual frontend connects the transformed spatial representations with gene expression and chromatin accessibility. The 3D views provide spatial context, while HDR and PF support direct comparison of structural changes and regulatory gene groups. MultiPASE also complements contact-matrix-based analysis by preserving and summarizing reconstructed genome geometry rather than replacing established contact-based measures~\cite{Rao2014A3dMap}.

The Supplementary Materials provide additional validation of the pipeline. The synthetic evaluation shows that HDR and PF preserve non-convex structure, the alignment ablation shows that alignment is necessary for meaningful comparison, and the chromosome-level analysis shows consistent relationships across abstraction levels. Together with the usability evaluation, these results support the stability and practical use of MultiPASE. MultiPASE is also designed as a general Python package that can be applied to 3D point-cloud collections beyond genome structures, including spatial transcriptomics and particle simulations. Its alignment, shape extraction, and comparison methods operate on generic point sets and do not depend on genomic annotations. On the visual side, screen space and 3D clutter can become limiting as the number of conditions, time points, or points increases, although filtering and interactive selection help users focus on relevant subsets.

Several limitations remain. MultiPASE inherits uncertainty from the input reconstruction, rigid alignment may not capture large non-rigid changes, and PCA initialization can become ambiguous for nearly symmetric point clouds. HDR and PF smoothing can reduce fine spatial detail, while the 2D projections collapse depth information. Gene positions are also limited by the 100 kb resolution of the reconstructed bead model. Future work will evaluate additional datasets and reconstruction pipelines and explore direct 3D shape abstraction and alternative rigid registration methods.

In conclusion, MultiPASE provides a computational pipeline and visual analytics tool for comparative analysis of reconstructed 3D genome structures. By combining rigid alignment with HDR and PF shape abstractions, it supports multi-scale comparison across conditions and time points while connecting spatial organization with gene expression and chromatin accessibility.

\section*{Acknowledgments}
% hidden
This work is supported by funds from the NIH NCI R01CA258827, NIH UG3 TR004501, NSF CNS-2320261, and the UIC Institute for Health Data Science Research. We thank the UIC Electronic Visualization Laboratory. This material is based upon work supported by the U.S. Department of Energy, Office of Science, through the Biological and Environmental Research (BER) and Advanced Scientific Computing Research (ASCR) programs under contract number 89233218CNA000001 to LANL (Triad National Security, LLC), awarded to S. R. Starkenburg and C. R. Steadman.

%-------------------------------------------------------------------------
% bibtex
\bibliographystyle{eg-alpha-doi} 
\bibliography{egbibsample}       

% biblatex with biber
% \printbibliography                

%-------------------------------------------------------------------------

\end{document}